\documentclass[namedreferences]{SolarPhysics}
 \usepackage[optionalrh,solaenum]{spr-sola-addons} 
\usepackage{graphicx}                    
\usepackage[usenames]{color}
\usepackage{url}                         

\newcommand{\ion}[2]{%
\relax\ifmmode
\ifx\testbx\f@series
{\mathbf{#1\,\mathsc{#2}}}\else
{\mathrm{#1\,\mathsc{#2}}}\fi
\else\textup{#1\,{\mdseries\textsc{#2}}}%
\fi}
\begin{document}

\begin{article}

\begin{opening}

\title{A Peculiar Velocity Pattern in and near the Leading Sunspot of NOAA 10781: Wave Refraction by Large-Scale Magnetic Fields?}

%
\author{C.~\surname{Beck}$^{1}$}

%
\runningauthor{C.~Beck}
\runningtitle{Wave Refraction by Large-Scale Magnetic Fields}

%
  \institute{$^{1}$ Instituto de Astrof\'{\i}sica de Canarias
                 }

\begin{abstract}
I report observations of unusually strong photospheric and chromospheric velocity oscillations in and near the leading sunspot of NOAA 10781 on 03 July 2005. I investigate an impinging wave as a possible origin of the velocity pattern, and the changes of the wave after the passage through the magnetic fields of the sunspot.

The wave pattern found consists of a wave with about 3 Mm {\em apparent} wavelength that propagates towards the sunspot. This wave seems to trigger oscillations inside the sunspot's umbra, which originate from a location inside the penumbra on the side of the impinging wave. The wavelength decreases and the velocity amplitude increases by an order of magnitude in the chromospheric layers inside the sunspot. On the side of the sunspot opposite to the impinging plane wave, circular wave fronts centered on the umbra are seen propagating away from the sunspot outside its outer white-light boundary. They lead to a peculiar ring structure around the sunspot, which is visible in both velocity and intensity maps. The fact that only weak photospheric velocity oscillations are seen in the umbra -- contrary to the chromosphere where they peak there -- highlights the necessity to include the upper solar atmosphere in calculations of wave propagation through spatially {\em and} vertically extended magnetic field concentrations like sunspots.
\end{abstract}

%

\end{opening}
\section{\label{sec_int}Introduction}
The effect of magnetic fields on the passage of (acoustic) waves is important for local helioseismology and time-distance analysis \cite{cameron+etal2008,moradi+cally2008,khomenko+etal2009}. The magnetic fields of sunspots can change the characteristic mode of the wave and selectively absorb certain frequencies only \cite{braun1995,cally2000}. Using the change of the wave pattern in and near sunspots, the properties of the solar velocity field even below the visible surface can be determined \cite{duvall+etal1996,haber2008}. Whereas in most of these publications the effect of the magnetic fields on impinging waves was studied, some groups also have investigated the fate of waves that are generated inside a sunspot \cite{khomenko+collados2006,olshevsky+etal2007,parchevsky+kosovichev2009}. \inlinecite{khomenko+collados2006} found for instance that not only the wave mode changes during the propagation inside inclined magnetic fields, but also the propagation direction as well, up to a downward reflection of upwards propagating waves (see also \opencite{khomenko+collados2009}). 
\begin{figure}  
\centerline{\resizebox{9.8cm}{!}{\hspace*{.2cm}\includegraphics{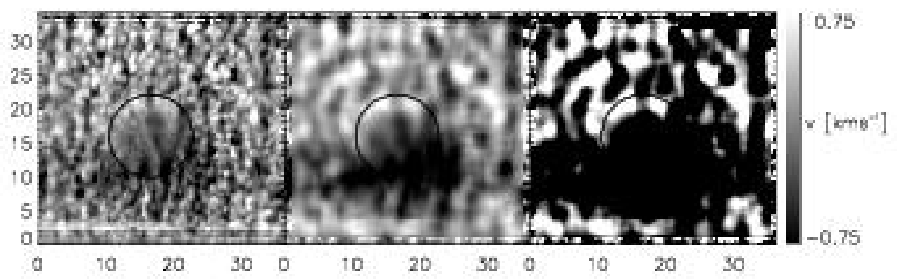}}}
\centerline{\resizebox{9.8cm}{!}{\hspace*{.2cm}\includegraphics{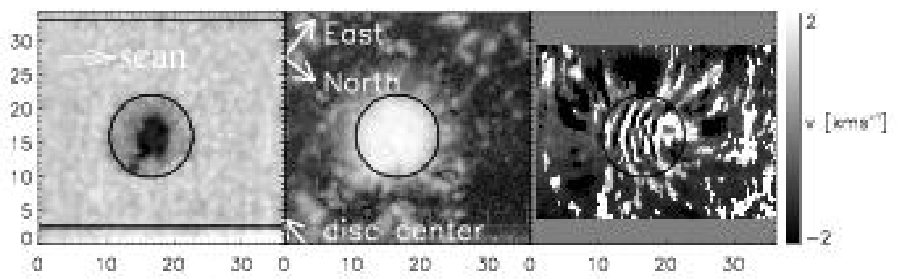}}}
\caption{Top row: LOS velocity of \ion{Ti}{i} at 630.37\thinspace nm in
  original resolution (left), smoothed by 15$\times$15 pixels (middle),
  and smoothed and contrast enhanced (right). The black circle
  marks the outer limit of the penumbra. Lower row: continuum intensity at
  630 nm, polarization degree, and LOS velocity of \ion{He}{i} at
  1083\thinspace nm. For \ion{He}{i} the displayed velocity range was
  doubled. Positive velocities correspond to blueshifts of the lines. The granulation contrast is about 3\%; the difference between the intensity in the umbra ($\approx$0.5 of I$_c$) and the quiet Sun ($\approx$I$_c$) strongly suppresses its visibility in the image. The various white arrows denote the scan direction from left to right, the direction towards disc center, and towards solar North and East, respectively. Tick marks are in Mm.\label{fig1}}
\end{figure}

The unusual velocity pattern near the leading sunspot of NOAA
10781 was actually detected only by chance. At the end of the usual reduction of data taken with the POlarimetric LIttrow Spectrograph \cite{beck+etal2005b},
maps of the line-of-sight (LOS) velocity of the observed spectral lines are
automatically created to be placed into an on-line overview archive (\url{http://www3.kis.uni-freiburg.de/~cbeck/POLIS_archive/POLIS_archivemain.html}).
The peculiarity of the observation in question went unnoticed up to that point, but a closer inspection of
any velocity map of the POLIS lines from the data set in the archive or
in the upper part of Figure 1 reveals at once the presence of bright and
dark rings, most prominent in the upper left half of the map. The rings
can be seen more clearly in the smoothed and contrast-enhanced
velocity map (top middle and left). Especially if one demagnifies
the image, as done automatically in the data archive, the visibility
is strongly enhanced. The ring pattern does not seem to be related
directly to photospheric magnetic fields; it is located partly
at places without a polarization signal (middle bottom panel of
Figure 1). The same pattern is visible also in the \ion{Si}{i} line at
1082.7 nm observed simultaneously with the \textit{ Tenerife Infrared Polarimeter}
(TIP: \opencite{martinezpillet+etal1999}; \opencite{collados+etal2007})
while the chromospheric \ion{He}{i} line at 1083 nm shows a markedly
different pattern with strong oscillations in the umbra. Taken together,
all spectral lines indicate unusually strong oscillations.

In the following section, I will discuss the observations in detail. I will
then use different approaches to determine the spatial and temporal properties
of the oscillations in the umbra and the surroundings of the sunspot in
Section \ref{ana}. I discuss a possible source of the observed  wave pattern in
Section \ref{disco}. The conclusions are presented in Section 5, while the Appendix A shows some numerical tests on the influence
  of the observation mode on the found wave properties.
\begin{table}
\caption{Spectral Lines Observed\label{speclines} }
\begin{tabular}{cccc}
\multicolumn{2}{c}{POLIS} & \multicolumn{2}{c}{TIP} \cr
\ion{Fe}{i} & 630.15 nm & \ion{Si}{i} & 1082.7 nm  \cr
\ion{Fe}{i} & 630.25 nm & \ion{He}{i} &  1083 nm\cr
\ion{Ti}{i} & 630.37 nm &  & \cr
\end{tabular}
\end{table}
\section{Observations and Data Reduction}
The leading sunspot of NOAA 10781 was observed on 03 July 2005 with the
two spectropolarimeters POLIS and TIP at the German \textit{Vacuum Tower
Telescope} (VTT) at Iza{\~n}a, Tenerife. The sunspot was located close to
disc center at a heliocentric angle of about 15$^\circ$. POLIS observed
the spectral range around 630\thinspace nm with three photospheric lines,
while TIP was set to record spectra around 1083\thinspace nm, including
a photospheric \ion{Si}{i} line and the chromospheric \ion{He}{i} line
at 1083 nm (see Table \ref{speclines}). Both instruments measured the
full Stokes vector in their respective ranges; the polarimetric
data was corrected for instrumental polarization as described by \inlinecite{schlichenmaier+collados2002} and \citeauthor{beck+etal2005a}
(\citeyear{beck+etal2005a,beck+etal2005b}). From
7:30 UT until 8:18 UT, a scan of 250 scan steps across the sunspot was done, with 12 seconds integration
time per scan step of 0.2$^{\prime\prime}$ step width. The slit width
corresponded to 0.5$^{\prime\prime}$ for POLIS and 0.36$^{\prime\prime}$
for TIP, respectively. The data were taken as part of a campaign under
the International Time Program, where all solar telescopes on the Canary
Island were run in coordination; the observations at La Palma ({\it Dutch
Open Telescope} (DOT) and {\it Swedish Solar Tower}), unfortunately, could
be started only about half an hour later due to the seeing conditions
there (\url{http://solarwww.mtk.nao.ac.jp/katsukaw/itp2005/}).

The TIP data were binned by a factor of two along the slit to a sampling of 0.35$^{\prime\prime}$ per pixel prior to evaluation to yield critical sampling of the diffraction limit of the 70-cm VTT at 1 $\mu$m. Maps of the LOS velocity were derived from the line-core positions of all spectral lines in the POLIS and TIP data, respectively. Positions in the POLIS spectra were measured relative to the telluric O$_2$ line at 630.20\thinspace nm to correct for the spectral curvature along the slit \cite{beck+etal2005b,rezaei+etal2006}. The rest wavelength for each line was then defined as the average line-core position inside the field of view (FOV). Umbral and penumbral points were excluded from the averaging, because the line-core position is not well defined due to the (partial) splitting of the lines. I applied no correction for convective blueshift or the moat flow to the rest wavelength. Positive velocities correspond to blueshifts of the line. In the following, I will mainly use the LOS velocity of the \ion{Ti}{i} line at 630.37 nm, because it is only weakly sensitive to magnetic fields and thus yields reliable line-core positions over the full FOV.
\begin{figure}
\centerline{\resizebox{9cm}{!}{\includegraphics{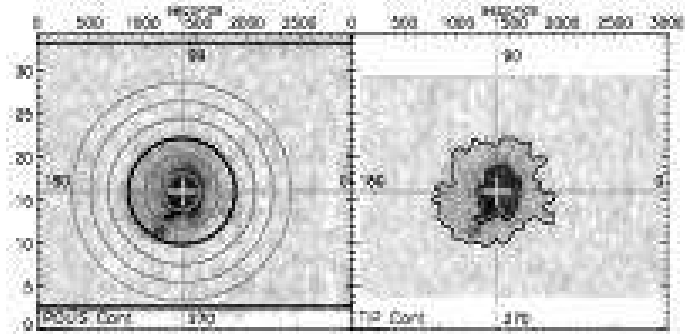}}}
\caption{Continuum intensity map of NOAA 10781 at 630\thinspace nm (left) and 1 $\mu$m (right). The spot center is marked with a white cross. The thick black ring marks the outer limit of the penumbra. The spatial scale is given in Mm; the upper $x$-axis gives the time in seconds. In the TIP continuum map, the boundaries of the penumbra in the POLIS map have been overlaid to visualize the alignment quality. \label{fig2}} 
\end{figure}

All 2D maps were finally re-sized to have square 0.2$^{\prime\prime}$-pixels. The FOVs of TIP and POLIS were aligned by correlating the continuum-intensity maps (Figure 2). The correlation yielded a displacement by 0.93$^{\prime\prime}$ between the two slits of POLIS and TIP in the scanning direction. To avoid another interpolation, I used a constant integer shift by five pixels of 0.2$^{\prime\prime}$ in the scanning direction to align the 2D maps; the shift was applied to all TIP maps. Due to the long integration time, this also means that the same spatial location inside the FOV was observed about 60 seconds {\em earlier} by TIP than by POLIS.
\section{Data Analysis and Results\label{ana}}
\subsection{Azimuthal Binning}
To verify the impression of the ring pattern in the velocity map, I took the values of velocity [$v$] and continuum intensity [$I_c$] along 100 rings that were centered on the umbra of the sunspot and whose radius was increased pixel by pixel (see Figure 2). This yielded the values of $v$ and $I_c$ on each ring in some azimuthal directions, where the azimuthal sampling is finer for the outer rings with their larger diameter. I then binned all quantities into azimuthal bins, whose size was 7.5$^\circ$ for the innermost 13 rings, 5$^\circ$ for rings 14 to 31, and finally 2.5$^\circ$ for rings 31 to 100. With the azimuthally binned values, the 2D maps can then be plotted linearly as a function of azimuth angle and distance from the spot center (Figure 3).
\begin{figure}
\centerline{\resizebox{11.5cm}{!}{\includegraphics{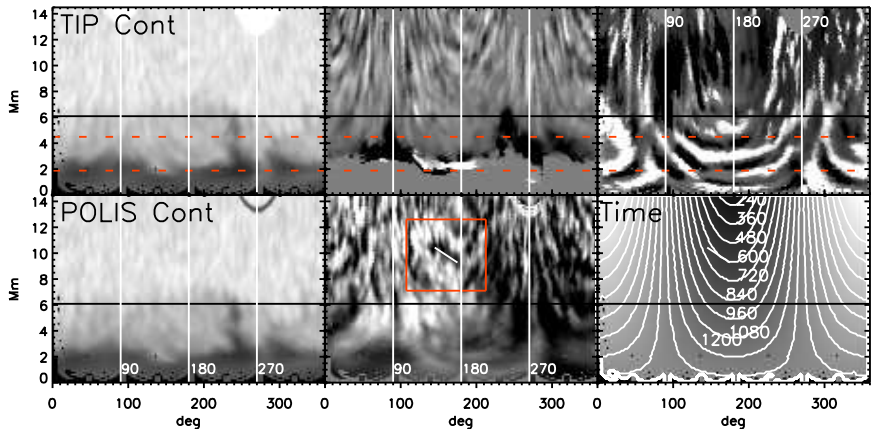}}}
\caption{Top row, left to right: maps of continuum intensity at 1\thinspace $\mu$m {\bf{\it vs.}}~azimuth angle and distance to the spot center, LOS velocity of \ion{Si}{i}, and LOS velocity of \ion{He}{i}. Bottom row, left to right: continuum intensity at 630.4\thinspace nm, LOS velocity of \ion{Ti}{i}, time since the start of the observation in seconds. The red rectangle outlines the location of the ring pattern in velocity; the inclined white line inside it marks one ridge. Each horizontal dashed line marks a decrease of the azimuthal bin size by 2.5$^\circ$. The black horizontal line at 6 Mm marks the outer end of the penumbra. In the map of the scanning time (bottom right), some isochrons have been overlaid. \label{fig3}}
\end{figure}

As the scan was performed with a long integration time per scan step, points
at equal radial distance from the spot center in different azimuthal
directions were actually observed at times that differ by several minutes. For
visualization of this effect, I also created a map of the time at which a
given spatial location was observed during the scan (bottom right panel
in Figure 3). Due to this temporal order in the observation, the evolution of
the wave pattern can be traced not only in space, but also to some extent in
time. Note that additionally TIP always observed the same spatial location 60
seconds earlier than POLIS.
\subsection{Phase speed, Velocity Amplitude and Wavelength of the Ring Pattern}
In the velocity map, the ring pattern around the spot is now transformed into a series of inclined ridges ({\it e.g.}, inside the red rectangle in Figure 3). The ridges are most prominent at an azimuth of about 140$^\circ$\,--\,180$^\circ$, the upper left half of the images in Figure 1. I chose one of the inclined ridges near an azimuth of 180$^\circ$ (white line) to derive the phase speed of the wave at this location. The velocity minimum is clearly defined; its distance from the spot center can then be plotted {\bf{\it vs.}}~time (left panel of Figure 4), using the information provided by the ``time'' map. A linear regression to the location of the ridge gave a phase velocity of about 8 km\,s$^{-1}$ which is comparable to the sound speed [$c_s$] in the photosphere. A second approach to determine the phase velocity is to use the temporal difference between the TIP and POLIS data for observing the same spatial location. The right panel of Figure 4 shows the intensities and LOS velocities on a cut from the sunspot center to the left, corresponding to an angle of 180$^\circ$ with the definition in Figure 2. The wave pattern in the velocity cut taken from the TIP data (thin grey line) is closer to the sunspot by about 0.7 Mm than in the POLIS data which have been taken 60 seconds later (thin black line), yielding a phase speed of about 12 km\,s$^{-1}$.
\begin{figure}
\centerline{\includegraphics{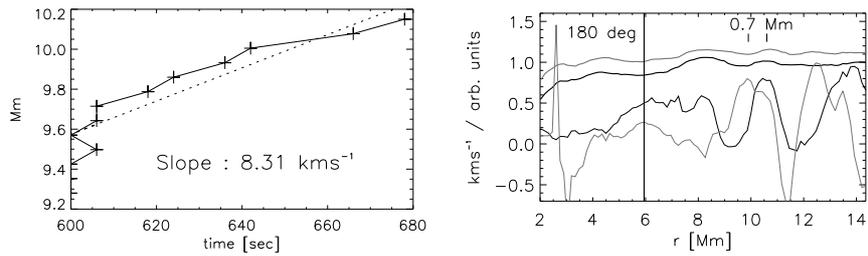}}
\caption{Left: Radial distance of the ridge marked in Figure 3 {\bf{\it vs.}}~time. The
  dashed line is a linear regression to the last eight data points. Right: cuts in intensity and velocity at 180$^\circ$ {\bf{\it vs.}}~distance to
    the sunspot center. Thick grey: infrared continuum intensity. Thin grey: LOS velocity of
    \ion{Si}{i} 1082.7\thinspace nm. Thick black: POLIS continuum
    intensity. Thin black: LOS velocity of \ion{Ti}{i} 630.37\thinspace
    nm. The IR data (grey lines) at the same spatial location is taken 60 seconds earlier.\label{fig5}}
\end{figure}
\begin{figure}
\centerline{\resizebox{6.cm}{!}{\includegraphics{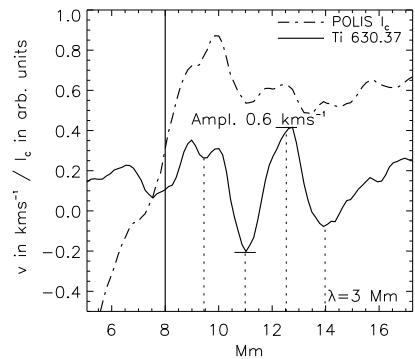}}}
\caption{Azimuthal average  from 130 to 180$^\circ$ of continuum intensity
  (dash-dotted) and LOS velocity of \ion{Ti}{i} 630.37\thinspace nm
  (solid).\label{finalfig1}} 
\end{figure}

The wavelength and amplitude of the velocity oscillation of the ring pattern
was determined from an azimuthal average from 130 to 180$^\circ$ (Figure 5). The propagation of the wave was removed by spatially
shifting the later scan steps before the azimuthal average. The velocity as
function of radial distance from the spot center then shows two clear maxima
and minima, whose separation corresponds to a wavelength of about 3 Mm. The
peak-to-peak amplitude of the velocity oscillation is about 0.6
km\,s$^{-1}$. Figures 4 and 5 also show that the oscillations are not restricted to velocity. The azimuthally averaged intensity in Figure 5 shows that the velocity extrema are correlated with intensity extrema; the intensity pattern, however, appears to be more diffuse, presumably due to the additional influence of the granulation pattern.

The derived wave properties (wavelength, phase speeds) are influenced by the specific type of the observation, scanning the slit with about twice the photospheric sound speed across the solar surface. Some numerical tests (see Appendix A) showed that the pattern still can be assumed to be caused by a propagating wave with properties comparable to the directly inferred values. 
\begin{figure}
\centerline{\includegraphics{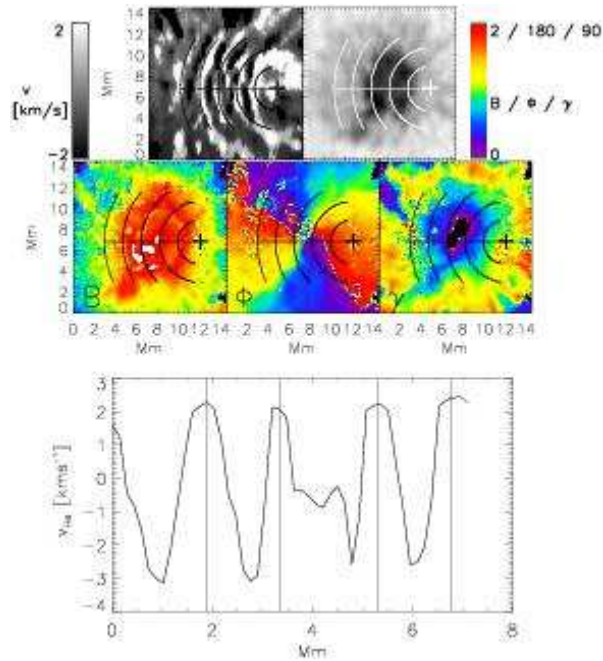}}
\caption{Close-up on the umbra. Top left: LOS velocity of \ion{He}{i}. Top right: continuum intensity at 1 $\mu$m. The oscillation pattern has been outlined by five rings with a spacing of 1.5 Mm. Middle row, left to right: field strength (kG), field azimuth (deg), field inclination (deg). The bottom panel shows the LOS velocity along the horizontal line through the center of the oscillation pattern.\label{umbrafig}}
\end{figure}
\subsection{Oscillations in the Umbra}
Inside the umbra, the chromospheric \ion{He}{i} line shows strong oscillations (Figures 1 or 3). Figure 6 displays a close-up of the umbra where the  pattern of the umbral oscillations is highlighted by five circles of increasing radius. If the center of these circles is actually the source of the oscillations, it is located outside the umbra in the middle of the penumbra to the right. To derive the properties of the magnetic field at that location, I did an inversion of the two \ion{Fe}{i} lines at 630 nm with the SIR code \cite{ruizcobo+deltoroiniesta1992}; the inversion setup was identical to that used in \inlinecite{beck2008}, {\it i.e.}, two magnetic components with constant properties along the LOS were used. The seeming source of the oscillation pattern is in no way outstanding in field strength [$B$], field azimuth [$\Phi$], or field inclination [$\gamma$] (middle panels of Figure 6). Only the component with the stronger magnetic field strength is shown, but also no peculiarity is seen there in the other component. The source, however, shows inclined magnetic fields with about 45$^\circ$ inclination. The wavelength of the umbral oscillations is about 1.45 Mm with a peak-to-peak amplitude of about 5 km\,s$^{-1}$, comparable to the values given by \inlinecite{tziotziou+etal2006} for umbral flashes. On a closer inspection of the bottom middle panel of Figure 3, one can see some of the umbral oscillations also in the photospheric \ion{Ti}{i} line, but with reduced amplitude; a similar reduction of photospheric oscillations in the umbra was reported by \inlinecite{kobanov+makarchik2004}. Figure 6 suggests that the umbral oscillations are uni-directional and propagate only to the left part of the FOV.
\section{Summary and Discussion \label{disco}}
The wave pattern inside the sunspot and in its immediate surroundings can be quantified as follows:
\begin{itemize}
\item[{\it i})] In the left part of the map (that is, at the beginning of the scan), waves with a wavelength of about 3 Mm and a peak-to-peak amplitude of 0.5\,--\,1 km\,s$^{-1}$ can be seen, which are receding from the spot with a phase speed of about 8 km\,s$^{-1}$. The waves are visible in the LOS velocity of all photospheric lines (\ion{Ti}{i} at 630.37\thinspace nm, \ion{Fe}{i} at 630.15, 630.25\thinspace nm, and \ion{Si}{i} at 1082.7\thinspace nm). In the upper left part of the map, the waves form a ring-like structure seemingly centered on the umbra of the spot. The wave pattern can also be identified in the continuum intensity. 
\item[{\it ii})] In the umbra of the spot, the chromospheric \ion{He}{i} line (and partly the photospheric \ion{Ti}{i} line) show strong oscillations with a wavelength of 1.5 Mm. The velocity amplitude in \ion{He}{i} is up to 5 km\,s$^{-1}$. These oscillations seem to originate in a part of the penumbra to the right that is not prominent in the magnetic field properties [$B,\Phi, \gamma$] in any way.
\end{itemize}
The numerical tests in Appendix A show that due to the scanning of the surface with a supersonically (relative to the sound speed in the photosphere) moving slit the wavelengths and phase speeds are at first only {\em apparent} quantities. The present case is even worse for deriving the real phase speeds or wavelengths than the commonly used observations with a two-dimensional FOV and high temporal cadence. The data are only a single snapshot, and the scanning introduces a Doppler effect that can only be fully taken into account if the propagation direction of the wave in all three spatial dimensions would be known. It seems, however, that both the shape of the velocity pattern with the closed phase fronts and the apparent phase speed outside the sunspot can be reproduced assuming an acoustic wave propagating with the photospheric sound speed of about 6\,--\,7 km\,s$^{-1}$ and a wavelength comparable to the observed one. Assuming a similar Doppler effect for the umbral oscillations, the real wavelength there will be larger than the apparent one of 1.5 Mm. 

The wave pattern differs significantly from oscillations inside sunspots (see, {\it e.g.}, \opencite{bogdan+judge2006}). A prominent feature there are the running
penumbral waves that were reported to have phase speeds of several tens of
km\,s$^{-1}$ ({\it e.g.}, \opencite{kobanov+etal2006}; \opencite{nagashima+etal2007}; \opencite{kobanov+etal2009}, and references therein). \inlinecite{rouppevandervoort+etal2003} and \inlinecite{bloomfield+etal2007} explained these strongly supersonic {\em apparent}
horizontal phase speeds with a similar argument as used in the appendix: the propagation of waves along magnetic field lines of varying inclination to the surface can create a visual pattern of a horizontally moving and expanding wave front, even if the wave propagation direction is far from horizontal. For the ring pattern outside the sunspot, a guiding of the wave by magnetic field lines is very unlikely, since the pattern is seen in the mainly field-free moat (see the polarization degree map in Figure 1). Even if the magnetic field lines of the penumbra can be followed over some distance beyond the white-light boundary of sunspots in the photosphere, as proven by the extension of the Evershed flow into the canopy area \cite{solanki+etal1994,rezaei+etal2006}, they have left the formation height of the photospheric spectral lines at the location of the outer part of the ring pattern. The numerical test with a spherical wave also suggests that the only slightly supersonic horizontal phase speed already results for a simple spherical wave in the field-free plasma without the need of guiding field lines. For the umbral oscillations, the observations offer the explicit option to investigate a possible guiding by magnetic fields as studied by \inlinecite{bloomfield+etal2007}. The spectropolarimetric data allows us to derive the magnetic field orientation inside the sunspot (Figure 6), and the information on both photospheric and chromospheric layers is available (see also \opencite{centeno+etal2006}).
\begin{figure}
\centerline{\resizebox{6.cm}{!}{\includegraphics{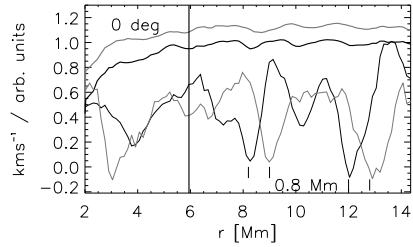}}}
\caption{Cuts in intensity and velocity at 0$^\circ$. Thick grey: infrared continuum intensity. Thin grey: LOS velocity of \ion{Si}{i} 1082.7\thinspace nm. Thick black: POLIS continuum intensity. Thin black: LOS velocity of \ion{Ti}{i} 630.37\thinspace nm. The IR data at the same spatial location is taken 60 seconds earlier.\label{finalfig}} 
\end{figure}

However, the main question still remains open: what is the source of the wave
pattern\,? A possible hint may be the shape of the oscillations in the umbra:
the assumed source is located in the right part of the penumbra. At the very
end of the scan, a plane wave can be seen that extends all along the slit and
that also shows up in the chromospheric He line as well (see
  Figure 1). Interestingly, this wave is moving {\em towards} the sunspot, not
receding from it. This is shown in Figure 7, which displays a cut at 0$^\circ$
(that is, from spot center to the right, compare to the right panel of Figure 4). At 0$^\circ$, the wave pattern is further away from the spot in the TIP data (grey lines)  taken 60 seconds earlier. The apparent phase velocity is again about 10\,--\,13 km\,s$^{-1}$. The wavelength of this impinging wave is about 3 Mm and the amplitude is up to 1 km\,s$^{-1}$ peak-to-peak, about twice as much as for the receding ring pattern.

There are two other possible sources for generating the wave pattern: a plage area close to the sunspot, where recurrent reconnection could trigger waves, or a flare (\opencite{kosovichev+zharkova1995}, \citeyear{kosovichev+zharkova1998}; \opencite{moradi+etal2007}). I think, however, that I can exclude both of these options. Comparing with the imaging data taken about half an hour later with the DOT (Figure 8), one finds that the plage area is located North-west of the sunspot (white rectangle), whereas the impinging wave comes from the North-east. According to an activity monitor (\url{http://www.lmsal.com/solarsoft/latest_events/}), no flares were related to NOAA 10781 on that day inside the time window of 8:00 UT$\pm$ 5 hours. Actually, no other active region of any kind was located in the direction of the impinging wave.
\begin{figure}
\centerline{\hspace*{.5cm}\resizebox{8.cm}{!}{\includegraphics{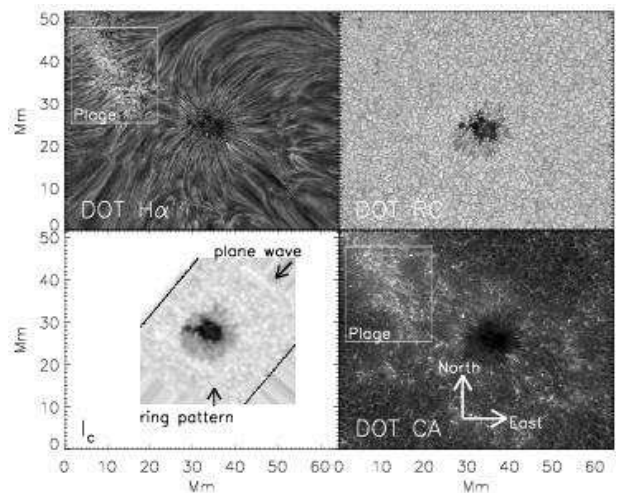}}}$ $\\
\caption{Images from the DOT, taken at about 09:00 UT. Clockwise,
    starting left top: H$_\alpha$ line core, red continuum, \ion{Ca}{ii}
  H. Left bottom: continuum intensity during the scan from 7:30 UT to 8:18 UT. \label{dot_img}}
\end{figure}

Thus, from the patterns found in the velocity and intensity maps, I would
explain the observed wave pattern around the leading spot of NOAA 10781 in the
following way: a plane wave ($\lambda \approx 3$ Mm, peak-to-peak velocity
amplitude $\approx 1$ km\,s$^{-1}$, phase speed $\approx$ sound speed, photospheric)
is propagating towards the sunspot from the right part of the map which
correspond to solar North-east. The wave is absorbed by the sunspot, and
triggers a series of oscillations in the penumbra that propagate through the
umbra and the left side of the penumbra ($\lambda \approx 1.5$ Mm, amplitude
$\approx 5$ km\,s$^{-1}$, chromospheric). After passing the outer penumbral
boundary, the wave pattern appears as circular wave fronts centered on the
umbra of the spot in the upper left half of the map ($\lambda \approx 3$ Mm,
amplitude $\approx 0.6$ km\,s$^{-1}$, phase speed $\approx$ sound speed,
photospheric). From the numerical tests in Appendix A, the receding wave  is presumably a spherical wave with a source outside of the photosphere. The source could be located either below or above the photosphere, since the problem is symmetric in $z$; the wave propagation direction has only to make a (small) angle to the vertical direction to produce the supersonic apparent phase speed from a propagation with sound speed.

This scenario, however, comes with a great caveat due to the very nature of
the observation by sequential scanning of the solar surface. Part of the ring
pattern and the umbral oscillations are co-temporal (seen in the same
slit-spectrum, at the middle of the scan), but they were observed something
like 10\,--\,20 min {\em earlier} than the plane wave that propagates towards the
sunspot. The causal relationship thus is far from secure; the plane wave seen
at the end of the scanning could also be fully unrelated to the ring pattern.
\section{Conclusions}
The observed velocity pattern is unique in the sense that none of the several
dozens sunspot observations with POLIS or TIP taken between 2003 and 2009 have
shown a similar pattern with a comparable velocity amplitude. The wave pattern is directly visible in a single snapshot without the necessity of using difference images, opposite to the photospheric signature of the flare-induced ``sunquakes'' that generally show smaller velocity amplitudes. If the causal relationship developed in the previous section is  correct, then the strong
magnetic fields of the sunspot have refracted a plane wave in the photosphere,
which presumably was a common part of the solar acoustic $p$-mode oscillations,
into a spherical acoustic wave, reducing the velocity amplitude but preserving
the wavelength. As inside the umbra mainly the chromospheric helium line shows
indications of strong oscillations, the wave should first have propagated to the upper atmosphere layers and then back to the photosphere
inside the sunspot. \inlinecite{khomenko+collados2006} found such an effect in their simulations, where a part of the wave energy was reflected downwards again,
which fits with the reduction of the photospheric velocity amplitude seen in
the observations here. This would imply that for modeling the wave propagation
of oscillations inside the strong magnetic fields of sunspots also
chromospheric atmosphere layers have to be included.

\begin{acks}
The VTT is operated by the Kiepenheuer-Institut f\"ur Sonnenphysik (KIS) at the Spanish Observatorio del Teide, operated by the Instituto de Astrof{\'is}ica de Canarias (IAC). POLIS has been a joint development of the KIS and the High Altitude Observatory, Boulder, Colorado. The Dutch Open Telescope is operated by Utrecht University at the Spanish Observatorio del Roque de los Muchachos of the IAC. C.B.~acknowledges partial support by the Spanish Ministry of Science and Innovation through project AYA 2007-63881. I thank the PI of the ITP campaign in 2005, Y.~Katsukawa, for the permission to use the data.
\end{acks}

\begin{appendix}
\section{Numerical Tests\label{num_test}}
 The observation mode of spatially scanning the surface with time
  complicates the interpretation of the phase velocity and the wavelength of
  the ring-like wave pattern. To investigate how the properties found could
  possibly be interpreted, I performed two simple numerical tests. 
\begin{figure}
\centerline{\includegraphics{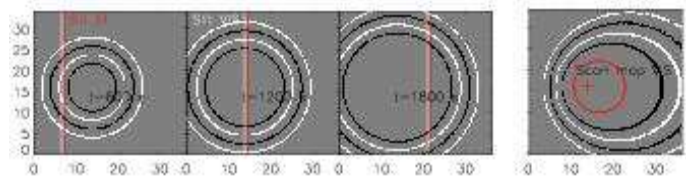}}
\caption{Left: circular radially propagating wave front at three instants of time. The locations of the two slits corresponding to the IR data (red) and the VIS data (white) are marked as well. Right: resulting velocity map  for the VIS slit after scanning. The red circle denotes the outer sunspot boundary as in Figure 1. The red cross marks the assumed origin of the wave. Tickmarks are in Mm.\label{sim1}}
\end{figure}
\subsection{Horizontal Circular Wave} I simulated a circular wave front that propagates radially outwards in the horizontal plane. The parameters of the wave were set to a wavelength of 4.2 Mm and a phase speed of 6 km\,s$^{-1}$. The origin of the wave was set 2.9 Mm to the left of the sunspot center (red cross in right panel of Figure 9). The wave propagation was started some time before the scanning to allow it to reach the left part of the FOV. The motivation for the choice of these parameters is discussed below. For simplicity, I considered only the location of the extreme amplitudes (set to $\pm 1$ in arbitrary units) of the wave for two full wavelengths. Three instants of the resulting circular wave front pattern at different times $n\times$600 seconds ($n$=1,2,3) are shown in the left panel of Figure 9. The size of the FOV shown corresponds to that of the observations. I then used two synthetic ``slits'' scanning the FOV while the wave was propagating, with properties identical to the observations: 12 seconds per scan step of 0.2$^{\prime\prime}$ and an 1$^{\prime\prime}$ separation between the slits, mimicking the separation between the TIP data in the infrared (IR) and the POLIS data in the visible (VIS). The location of the two slits at the different times is marked in Figure 9 as well (red and white vertical lines). The right panel shows the resulting velocity map as seen through the VIS slit. The effect of the sequential scanning modifies the shape of the circular wave front significantly, it actually corresponds to a strong Doppler effect between the (supersonically) moving slit and the propagating wave. 

I then made the same cut as in the right panel of Figure 4  from the sunspot center to the left in both the synthetic VIS and IR scan map (Figure 10). The cut actually compares fairly well to Figure 4 in the wavelength of about 2.8 Mm. This {\em apparent} wavelength [$\lambda^\prime$] can actually be derived exactly as $\lambda^\prime = \lambda \cdot \frac{v_{\rm slit}}{v_{\rm rel}}= 2.8$ Mm, where $v_{\rm slit} = 12$ km\,s$^{-1}$ and $v_{\rm rel}=$ 18 km\,s$^{-1}$. The separation of 0.29 Mm between the wave fronts in the VIS and IR maps, however, does not match the 0.7 Mm of the observations. 
\begin{figure}
\centerline{\resizebox{6.cm}{!}{\includegraphics{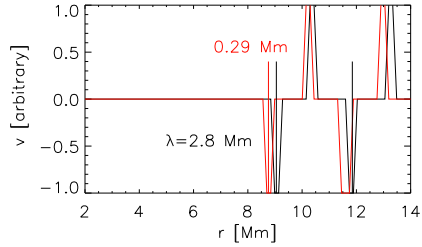}}}
\caption{Cuts in the velocity maps corresponding to the IR (red) and VIS slit (black) from the center of the sunspot to the left for a horizontally propagating circular wave.\label{sim2}}
\end{figure}

The parameters used for the simulated wave differ from the numbers derived from the observations (phase speed $>$ 8 km\,s$^{-1}$, wavelength 3 Mm), and the origin is not centered on the sunspot. It turned out to be impossible without these modifications to reproduce the shape of the phase fronts in the observations. On closer inspection, Figure 1 reveals that all phase fronts close {\em inside} the observed FOV, even the outermost ridge of positive velocity (white color) to the left. If the phase speed of the simulated wave was increased only slightly to above 6 km\,s$^{-1}$, this feature could not be reproduced in the simulation, a part of the phase front left the observed FOV. The effect critically depended on the assumed horizontal phase speed; in principle the motion of the slit has to be significantly faster than the propagation of the wave to partly ``freeze in'' the originally circular shape and to overtake the wave front in the later part of the scan. The phase fronts also did not close anymore when the origin of the wave was not displaced towards the left. The curvature of the individual phase fronts decreases rapidly with increasing radius; if the radius of the phase fronts at $x\approx$ 3\,--\,7 Mm was only slightly larger, the first two phase fronts left the FOV. The finally assumed location of the source is, however, still well within the sunspot. 
\subsection{Spherical Wave with an Origin below the Surface}
Since the observations were made close to disc center and the ring-pattern is seen in the LOS velocity maps, a pure horizontal propagation of the wave is actually unlikely since the LOS is almost perpendicular to the surface. A LOS velocity signal due to a horizontally propagating transverse magnetic wave mode can presumably be excluded. The polarization signal of the spectral lines comes from layers close to where the intensity profile near the line core forms, and in the moat the photospheric lines indicate basically a field-free atmosphere with only isolated magnetic patches without the necessary radial alignment (bottom middle panel of Figure 1). As second test, I thus considered a spherical wave starting from an origin below the surface. I used the same phase speed of 6 km\,s$^{-1}$, but a wavelength of 3.2 Mm only. The origin of the wave was set to the same location in the horizontal plane, but displaced to $z$= -15 Mm in the vertical direction. The corresponding motion of the phase front in the photosphere with its comparably small vertical extent was then determined by the intersection of the spherical wave with a horizontal plane at $z$= 0 Mm. Figures 11 and 12 then show exactly the same plots as for the case of a horizontal wave. Here the spatial pattern of the closed phase fronts is reproduced slightly worse, but the separation between the wave fronts in the VIS and IR maps of 0.5 Mm is significantly closer to the observations. In this case, the {\em apparent} phase speed in the horizontal plane of 8.3 km\,s$^{-1}$ is larger than the propagation speed of the spherical wave due to the angle between the horizontal plane and the wave propagation direction.  
\begin{figure}
\centerline{\includegraphics{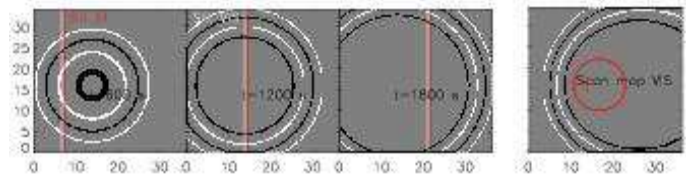}}
\caption{Left: spherical wave with an origin below the surface at three instants of time. The locations of the two slits corresponding to the IR data (red) and the VIS data (white) are marked as well. Right: resulting velocity map  for the VIS slit after scanning (compare with Figure 9). The red circle denotes the outer sunspot boundary as in Figure 1. The red cross marks the assumed origin of the wave. Tickmarks are in Mm. }
\end{figure}
\begin{figure}
\centerline{\resizebox{6.cm}{!}{\includegraphics{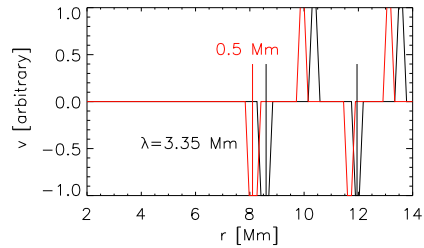}}}
\caption{Cuts in the velocity maps corresponding to the IR (red) and VIS slit (black) from the center of the sunspot to the left for the spherical wave (compare with Figure 10).\label{sim4}}
\end{figure}

The results of the numerical tests suggest that it should be possible to reproduce the ring pattern seen in the LOS velocity outside the sunspot with a propagating spherical wave with a suitable combination of phase speed, wavelength, location of the source, and starting time. Modifying the height of the wave source or the real phase speed changes the apparent phase speed in the horizontal plane, the distance to the wave source or the time since the start of the propagation modifies the curvature of the wave fronts, and the finally observed apparent wavelength can be matched exactly with a slight modification of the assumed initial wavelength. For a consistent reproduction, one would, however, also have to include the variation of the wave propagation speed in the solar atmosphere that changes with height and in the presence of magnetic fields ({\it e.g.}, \opencite{mihalas+mihalas1984}; \opencite{bodo+etal2000}; \citeyear{bodo+etal2001}).
\end{appendix}

\bibliographystyle{spr-mp-sola-cnd} 
\bibliography{references} 

\end{article} 
\end{document}